# 人口生命表与预期寿命计算新方法

**摘要**：现有生命表编制方法需要先计算年龄别死亡率，不仅计算步骤多而复杂，而且引入近似，从而存在误差。本文重新定义了生命表中死亡概率为某一年度出生的人群在其后某一时间段的平均死亡概率，以此定义为基础编制生命表，得到人均预期寿命，所代表的含义与传统生命表相同。使用日本人口数据验证了本文方法，结果显示，它与传统方法得到的出生婴儿预期寿命是很一致的，最大相对差别不超过千分之一，平均相对差别不到万分之三。本文所述生命表编制理论和方法简明易懂，所需要的统计数据也很容易获得，得到的各项数据都很准确可靠，而且容易计算，是一项有较大应用价值的人口学方法。
**关键词**：死亡概率；预期寿命；生命表；插值

A new method for life table and life expectancy calculation
Abstracts: The existing life table method needs to calculate the age-specific mortality first, not only has too many and complicated calculation steps, but also introduces the multiple approximation to bring error. This paper redefines the probability of death for the life table as the average probability of death of a group of people born in a certain period at a later time. Based on this definition, a new method for the life table is proposed to obtain the life expectancy, which has the same meaning to that from the traditional life table. Using the Japanese population data to verify the method, the results show that it is consistent with the life expectancy of the birth of the baby, the maximum relative difference is no more than 0.1%, and average relative difference is less than 0.03%. The theory and method of life table described in this paper are simple and easy to understand. The needed data are easy to obtained from statistics, and the calculation is easy, the results obtained are accurate and reliable. It should be a very valuable demographic method for application.

## 一、引言

人口生命表又称死亡表、寿命表[1]，是反映同时期出生的一批人随着年龄增长陆续死亡的整个生命过程的一种统计表格。在 Preston 所著《人口学》(demography)中，大部分内容都与生命表中概念和数据相关[2]，生命表是西方人口学的基础和核心，也是主导技术[3]。人们需要根据人口统计数据，按照生命表方法，获得人口的主要基础数据，用于下一步研究和分析。建立人口寿命表，对于研究人口再生产过程的生命现象有极重要的意义，能够计算今后要出生的一代人，按现有的社会条件，预期平均寿命是多大。

学者们提出了多种计算方法[4]，能够成功地根据统计数据计算得到死亡概率 $_nq_x$，从而构建人口生命表。早在 1939 年 Reed 和 Merrell 提出了计算死亡概率 q 的指数型经验关系式[5]：

$$_nq_x = 1 - \exp[-_nm_x * (a + b *_n m_x)] \tag{1}$$

其中 a 和 b 是作者根据美国人口统计数据确定的与 x 相关的经验常数，$_nm_x$ 是当年 x 到 x+n 岁死亡人口数 $_nD_x$ 与年内 x 到 x+n 岁年龄的平均人口的比值，为计算方便，常用年中时刻 x 到 x+n 岁年龄的人口数 $_nP_x$ 代替平均人口[6]。1943 年 Greville 推导了一个更为复杂的指数型近似计算式[7]：

$$_nq_x = 1 - \exp[-n * {_nm_x} - \frac{n^2}{12}({_nm_x})^2 * (ln_nm_x)'] \tag{2}$$

其中$(ln_nm_x)'$变化较小，在 0.08 到 0.11 之间。当 n=5 时，等于 0.096。该式与(1)式在形式上是相同的，只是选取的经验常数不同。

1975 年 Keyfitz 和 frauenthal 假定年龄分布和年龄别死亡率都是独立线性变化的，发展了适用性较普遍的指数型计算式[8]：

$$_nq_x = \exp[-{_nm_x} + \frac{n}{48\,_nP_x}({_nP_{x+n}} - {_nP_{x-n}}) * ({_nm_{x+n}} - {_nm_{x-n}})] \tag{3}$$

其中$_nP_x$是年中年龄在 x 到 x+n 岁之间的人口，该式不需要引入经验参数，但在年龄结构不稳定时，估算结果并不比其他方法好[2]。

1961 年蒋庆琅发展的经验计算式，因世界卫生组织推荐，成为比较常用的方法，其计算式[9]为：

$$_nq_x = \frac{n\,_nm_x}{1+(1-{_na_x})*n*{_nm_x}} \tag{4}$$

其中$_na_x$是 x 到 x+n 年龄组死亡人口的平均存活年数比例，又称终寿区间成数。通常$a_0$变化较大，蒋根据$m_0$推荐了所对应的$a_0$经验值，其他$a_x$变化较小，在没有经验数据时，可取值 0.5。

Berkson 公式则考虑了迁徙的影响，直接根据统计数据计算死亡概率[10]：

$$q_x = 1 + \frac{D_x}{2P_x} - \frac{S_x}{2P_x} - \sqrt{(1+\frac{D_x}{2P_x} - \frac{S_x}{2P_x})^2 - 2\frac{D_x}{P_x}} \tag{5}$$

$S_x$是迁出人数。Elveback 公式则同样考虑迁徙影响[11]：

$$q_x = 1 - (\frac{P_{x+1}}{P_x})\frac{D_x}{D_x + S_x} \tag{6}$$

当忽略迁移影响时，（5）和（6）式都变成：

$$q_x = \frac{D_x}{P_x} \tag{7}$$

与蒋庆琅公式相比，它们都在分母中忽略了死亡人数，从而高估了死亡概率。

London 在假设死亡人口对$P_x$影响可以忽略时，提供的计算公式为[12]：

$$q_x = \frac{D_x}{(p_x(t)+p_x(t+1))/2 + D_x/2} \tag{8}$$

此公式中$p_x(t)$和$p_x(t+1)$分别是年初和年末年龄是 x 的人口，它们的平均值可近似地看成是年中年龄为 x 的人口$P_x$。美国 CDC 直接根据统计资料计算死亡概率，其计算公式可根据蒋公式推导得到，如下[13]：

$$q_x = \frac{D_x}{P_x + D_x/2} \tag{9}$$

$P_x$是通过多个年龄的数据插值得到的。此外，还有很多其他方法和计算公式，如 King 方法，Wiesler 方法和 Sirken 方法等[9]。

然而，上述所有这些公式和方法都是在引入与实际存在差异的近似后推导出来的，虽然误差通常比较小，但实际上不可避免；而且它们的计算过程都很复杂。最近我们提出了新的直接计算死亡概率的公式[14]，

$$q_x = \frac{D_x/p * [1+(1-2f_x)*\delta/p - f_x*(1-f_x)D_x/p]}{1 - \delta^2/p^2} \tag{10}$$

其中 $f_x$ 是分离因子，p 和 $\delta$ 可按下式计算：

$$p=[p_x +p_{x+1}+D_x+f_{x-1}*D_{x-1}]/2 \tag{11}$$

$$\delta =p-(f_x*D_x+p_{x+1}) \tag{12}$$

该式虽然没有引入近似，可准确计算死亡概率，但需要分离因子 $f_x$。由于不能根据统计数据直接计算 $f_x$，我们提出插值迭代法，可根据统计数据近似计算 $f_x$，但仍出存在误差，而且计算过程很复杂。

本文从统计数据出发，重新定义生命表中死亡概率的物理意义，其基本出发点是以各年度出生人口在同一时期，处于不同年龄段的人口数和死亡数为依据，计算各人群的死亡概率和不同年龄的预期寿命，从而提出了新的生命表构建和人均预期寿命的计算方法。

二、生命表编制新方法
1、死亡概念新定义与计算方法

在生命表构建中，经常使用的人口统计数据是某一时刻各年龄段的人口数量以及此前一年以死亡人口年龄分组的死亡人口数量数据。例如，我国最近几次人口普查，以及人口抽查数据，都是如此。分年龄死亡人口通常是两个年龄段死亡人口之和，却没有给出每个年龄组人口的死亡人口，给死亡率计算带来了很多困难。例如，婴儿死亡人口既包括当年出生(0-1 岁年龄组)的死亡人口，也包括上一年出生，1-2 岁年龄组的部分死亡人口，按照定义，需要分别计算它们对婴儿死亡率的贡献，但这两个年龄段所贡献的死亡人口数是未知的，无法直接计算,只能近似处理。

下面更一般地分析依据统计数据计算得到的死亡概率 $q_x$ 的实际物理意义。按照死亡概率定义，通常 $q_x$ 是指某一时刻年龄为 x 的人群在今后 1 年的死亡概率，而实际统计资料是某一时刻有一定年龄范围人群的人口数量和此前一年死亡人口数量数据，因此，实际计算的死亡概率是相关人群在一年时间内死亡概率的平均值，由于编制生命表的基本假设是死亡概率不随时间变化，该平均结果可代表死亡概率。相关人群的共同特征是死亡年龄在 x 到 x+1 之间，它包括统计数据中两个年龄组，其一是年初年龄为 x-1 到 x 岁年龄组，其平均年龄为 x-0.5，其到达 x 岁年龄后的死亡，才与 $q_x$ 相关，他们在此年度经过的平均时间为半年；其二是年初年龄为 x 到 x+1 岁，其平均年龄为 x+0.5，但仍有很多死亡者的死亡年龄小于 x+1，在此年度经过的与 $q_x$ 相关的平均时间，同样为半年，需要分别计算，再汇总得到 $q_x$。

类似地，生命表中的死亡概率 $_nq_x$ 是指特定时刻活到年龄为 x 的人群，在此后 n 年的死亡概率[15]。在现有生命表编制方法中，我们需要根据统计数据，近似计算中心死亡率 $_nm_x$，再根据中心死亡率，近似地转换得到平均死亡概率 $_nq_x$。由于多次引入近似性的假设，不可避免地带来误差；加上计算过程复杂，不仅费时费力，而且容易出错，带来更大的问题。

本文定义的死亡概率 $_nQ_x$ 是指某一起始时刻年龄为 x-1 到 x 岁人群，在此后 n 年的死亡概率的平均值。与 $_nq_x$ 定义不同的地方在于，死亡年龄的范围不同，例如，n=1 时，传统定义的死亡概率计算中，死亡者死亡时年龄范围限定在 x 到 x+1 岁，本文定义的 $Q_x$ 的死亡者年龄范围则为 x-1 到 x+1。$_nq_x$ 定义的起始年龄和时间跨度是固定的，变化的是起始时间，是对起始时间的平均值。两者定义的时间长度也有所不同，对 $_nq_x$ 来说，讨论的时间长度总是 n 年，但对 $_nQ_x$ 来说，当 x=0 时，考察人群的时间长度也是变化的，从 n-1 年到 n 年变化，平均时间长度是 n-0.5 年。如果统计的分年龄死亡人口是对应年龄段的死亡人口，也就是根据统计标准时间计算死亡者年龄来分组，则根据本文定义的死亡概率：

$$_nQ_x=n \text{ 年死亡人口数}/(\text{期末人口数}+n \text{ 年死亡人口数}) \tag{13}$$

n 年死亡人口数都是 x 年龄组在 n 年内的死亡人口。当 n=1 时，可以根据人口统计资料直接计算得到，不需要引入任何假设。与传统方法相比，本方法根据统计数据，对死亡人口分组不同于传统方法，是根据年初年龄进行分组，而不是传统方法根据死亡年龄进行分组，

其优点在于，死亡人口分组与人口分组是一致的，从而可以式(13)精确计算死亡概率 Q。

2、传统死亡概念 $q_x$ 计算

本文使用的日本死亡数据给出了两个年龄段对死亡人口的分别贡献，可以准确计算 $q_x$，从而可以按照传统方法构建生命表，可用于验证本文提出的新方法，这里介绍使用该数据集计算 $q_x$ 方法。如果使用 $D_x$ 代表统计给出的年内 x 到 x+1 岁的死亡人口数量，他们中年初年龄在 x 到 x+1 岁人口为 $D_{x-}$，年末统计仍然存活的人口数为 $p_{x+1}$；年初年龄为 x-1 到 x 岁，死亡人口为 $D_{x+}$，在年末统计出的存活人口数为 $p_x$。则他们在年内死亡时，年龄为 x 到 x+1 岁的死亡概率则为：

$q_{x2}=D_{x+}/[\ D_{x+}+p_{x+1}]$;   (14)

而年初年龄在 x-1 到 x 岁人口的死亡，在年内死亡时，年龄为 x 到 x+1 岁的死亡概率则为：

$q_{x1}=D_{x-}/\ [D_{x-}+D_{(x-1)+}+p_x]$;   (15)

当 x=0 时，上式则变成：

$q_{x1}=D_{x-}/\ [D_{x-}+p_x]$;   (16)

则 x 到 x+1 年龄组的死亡概率[16]：

$q_x=1-(1-q_{x1})*(1-q_{x2})=q_{x1}+q_{x2}-q_{x1}*q_{x2}$;   (17)

3、生命表编制

使用传统定义的死亡概念 q 和本文定义的死亡概念 Q，分别按照传统方法编制生命表[2]，主要步骤如下：

1) 首先选择生命表基数 $l_0$=100,000;
2) 计算 $l_{x+n}=l_x*(1-_nq_x)$，对于本文方法，$l_{x+n}=l_x*(1-_nQ_x)$。
3) 计算 $_nd_x=l_x-l_{x+n}$
4) 计算生存人年数 $L_{x+n}$

有了死亡概率，很容易计算生存人数与死亡人数，但是，在计算生存人年数时，却需要引入近似，通常是根据历史数据总结的经验公式计算，如下：

$_nL_x=n*l_{x+n} + _na_{x+n}*_nd_{x+n}$   (18)

$_na_{x+n}$ 代表了死亡人群在 x 到 x+n 年龄段时的平均生存时间，人们总结了一些经验数据，在实际计算时，在大多数情况下，都取 n/2。

本文同时采用插值法计算了累积死亡人口数据，在获得 $l_x$ 数据集后，可以插值得到 $l_x$ 和 $l_{x+n}$ 之间的 $l_x$ 数据，从而可以得到 $L_x$。这种方法，不需依赖历史数据，仅根据现有数据集，因而不存在历史数据与现实之间的差异带来的误差，是通用性更好的方法。在计算 $L_x$ 时，首先通过插值法给出均匀分布在 x 到 x+n 年龄段的 m 个年龄的存活人口 $l_x$ 和累积死亡人口数，从而计算得到死亡人口的在该年度的平均存活时间 $a_x$。

5) 计算 $T_x=\sum_{a=x}^{\infty} {_nL_a}$   (19)
6) 计算平均预期寿命 $e_x=T_x/l_x$   (20)

三、本生命表编制方法的验证

日本统计部门提供了 2009-2017 年各年度分年龄死亡人口在两个出生年份的统计数据，部分数据如下：

表 1、部分日本 2010 年分年龄死亡人口及其出生年份的统计数据

| 出生年份 | 2010 | 2009 | 2009 | 2008 | 2008 | 2007 |
|---|---|---|---|---|---|---|
| 死亡年龄 | 0 | 0 | 1 | 1 | 2 | 2 |
| 总 | 1923 | 558 | 214 | 187 | 108 | 98 |
| 男 | 1096 | 306 | 126 | 100 | 56 | 56 |
| 女 | 827 | 252 | 88 | 87 | 52 | 42 |

这些年份中只有 2010 和 2015 年有人口普查给出的分年龄人口分布数据，使用这两个年份数据，可使用式(14)—(17)准确计算各年龄的死亡概率 $q_x$。由于原始的死亡人数和年龄分布数据在时间上不一致，以及存在未知年龄人口及死亡人口，本文根据出生人口和死亡人口的月份分布以及未知年龄数据，校订了数据，使两者在统计时间上一致。

另一方面，则可以根据该组数据，按照年初年龄或出生年份，对死亡人口重新分组，如下表：

表 2、日本 2010 年死亡人口按年初年龄分组的统计数据

| 年初年龄(岁) | 0 | 0--1 | 1--2 | 2--3 | 3--4 | 4--5 |
|---|---|---|---|---|---|---|
| 年初平均年龄(岁) | 0 | 0.5 | 1.5 | 2.5 | 3.5 | 4.5 |
| 总 | 1923 | 772 | 295 | 193 | 159 | 134 |
| 男 | 1096 | 432 | 156 | 103 | 104 | 75 |
| 女 | 827 | 340 | 139 | 90 | 55 | 59 |

使用表 2 数据和分年龄人口数据，就可以根据公式(13)计算本文定义的死亡概率 $Q_x$。图 1 是用两种方法计算得到的两种死亡概率，由于两种方法计算得到的死亡概率的分组不同，不能直接比较。传统方法得到的死亡概率是起始年龄为整数的年龄组死亡概率，而新方法得到的死亡概率，除出生婴儿外，都是平均起始年龄为 n+0.5 岁年龄组，也是同一年出生人口的平均死亡概率，两者在大部分年龄都组成了很好的曲线，表明结果是相当一致的。只有在几个极高年龄组，由于统计人数很少，两者统计对象不同，导致较大差异。

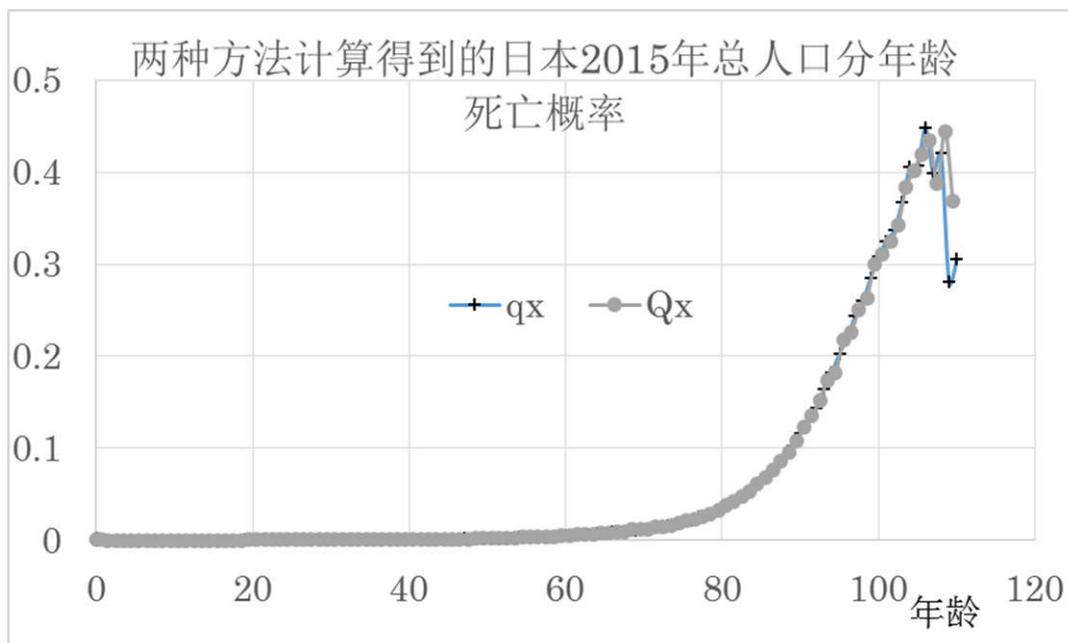

图 1 使用传统法和新方法分别计算得到的日本 2015 年总人口分年龄死亡概率

图 2 是使用两种方法计算得到的日本 2015 年人口的分年龄预期寿命，虽然两种方法给出的预期寿命，除 0 岁外，都是不同年龄，但它们组成了很好的曲线，表明两者是有较好的一致性的。更重要的证据来自出生婴儿的预期寿命，两者给出的预测结果非常相近，如表 3 所示，相对误差不到千分之一。

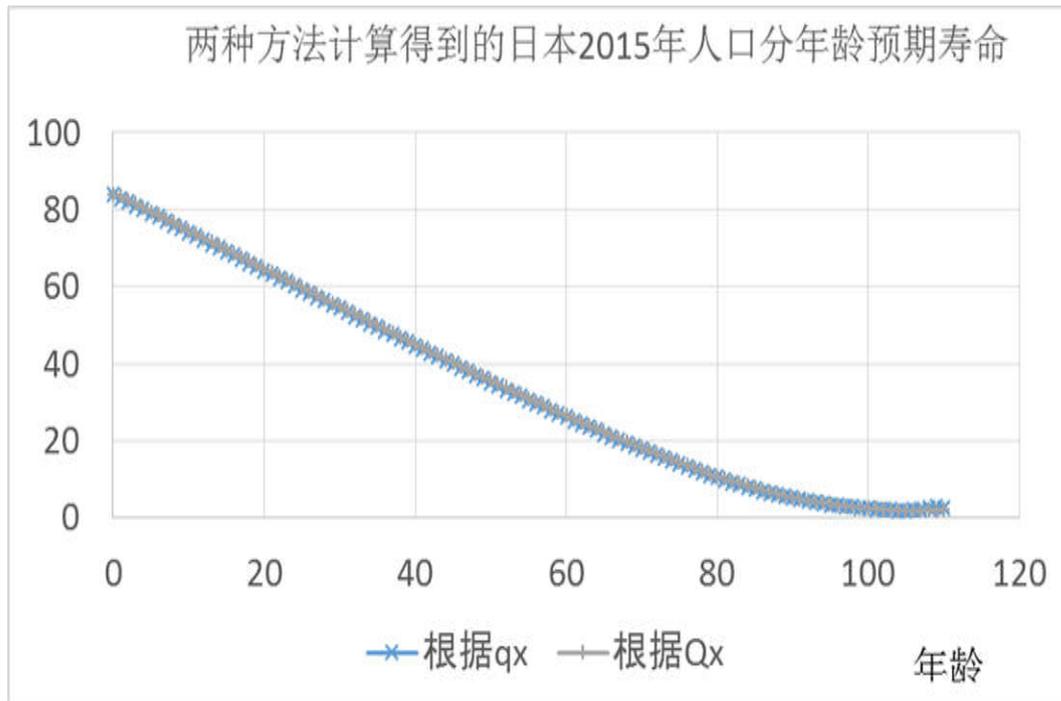

图 2 使用两种方法计算得到的日本 2015 年总人口分年龄预期寿命

表 3 给出了使用两种方法计算得到的 2010 和 2015 年日本出生婴儿预期寿命，从表中可以看出，两种方法所得出生婴儿预期寿命是非常一致的，最大相对差别不超过 0.1%，平均相对差别仅有 0.028%，不到万分之三。因此，本文发展的基于新定义的死亡概率和生命表编制方法，与传统方法是非常一致的。不同的插值法，如线性插值，三次样条插值以及三次方插值，给出的结果差异更小，相对差别仅在百万分之十以下，就不一一报告了。另插值法和固定 $a_x=0.5$ 法所得结果也是基本一致的，其原因在于，虽然 $a_0$ 通常比 0.5 小很多，但日本 $q_0$ 很小，使得 $a_0$ 的取值对出生婴儿预期寿命的影响很小；其他 $a_x$ 与 0.5 的差异较小，使得两种方法所得结果的差异很小。由于缺少能够准确计算死亡概率 $q_x$ 的统计数据，也就难以验算 $q_0$ 较大时，采用固定 $a_0$ 对生命表编制和出生婴儿预期寿命的一般影响了。

表 3  两种方法计算得到的 2010 和 2015 年日本出生婴儿预期寿命

|  |  | 2010 | | | 2015 | | |
|---|---|---|---|---|---|---|---|
|  |  | 总 | 男 | 女 | 总 | 男 | 女 |
| 插值法 | 根据 qx | 83.0383 | 79.5295 | 86.3825 | 83.7727 | 80.5771 | 86.8287 |
|  | 根据 Qx | 82.9913 | 79.5161 | 86.3102 | 83.7745 | 80.5835 | 86.8297 |
|  | 相对误差% | −0.057% | −0.017% | −0.084% | 0.002% | 0.008% | 0.001% |
| 固定 $a_x$ | 根据 qx | 83.0383 | 79.5295 | 86.3825 | 83.7727 | 80.5771 | 86.8287 |
|  | 根据 Qx | 82.9913 | 79.5161 | 86.3102 | 83.7745 | 80.5835 | 86.8297 |
|  | 相对误差% | −0.057% | −0.017% | −0.084% | 0.002% | 0.008% | 0.001% |

四、讨论和结论

虽然编制生命表已经有很长历史了，但目前文献报道的方法仍然存在明显的缺陷，由于现有方法需要数据较多，而人口统计仅提供了少量数据，缺少了必要的数据，实际计算时，常常引入了多种近似，使得计算过程引入了多种误差，也使得计算结果的可靠性降低了。常用的算法有三项根本的缺陷[15]，其一是定义的年龄别死亡率单位是 1/年[1]，和死亡概率单位就不匹配，不应用来计算死亡概率；其二是计算出生婴儿死亡概率，因缺少必要数据，常常

忽略不同年份出生婴儿数量变化，使用的出生数据往往仅包括婴儿在当年出生数据[17]；其三是计算分年龄死亡概率，往往使用的人口数据和死亡数据是不对应的，对于高年龄段人口，往往带来误差[2]。

本文提出了新的死亡概率概念，在此基础上提出了生命表编制方法与人均预期寿命计算方法，其基本的依据在于重新定义生命表死亡率含义，将其定义为某一年度出生的人群在其后另一时间段的平均死亡概率。

根据本文重新定义的死亡概率，同样可以构建生命表，计算平均预期寿命，得到的平均预期寿命是某一年度出生人口的平均预期寿命，这与依据现有生命表计算得到的平均预期寿命的含义一致。其计算方法的差别在于，依据本文计算得到的 $Q_0$ 是 0 岁婴儿经过 0.5 年死亡概率的平均值，不是传统定义的 0 岁婴儿经过 1 年的死亡概率平均值，因而 $Q_0$ 必然小于 $q_0$，但对 $Q_1$ 来说，所阐述的人群包括出生仅有 1 天到 365 天的婴儿，平均年龄仅为 0.5 岁，小于现有定义的从 1 岁开始计算，等于比所阐述人群小半岁，由于婴儿在出生后第一年的死亡概率通常大于其后几年的死亡概率，其死亡概率 $Q_1$ 必然大于 $q_1$。到高年龄段，由于较大年龄死亡率较高，$q_x$ 的年龄大于 $Q_x$，所以，$q_x$ 多大于 $Q_x$。

由于两者在描述 0 岁以上人群年龄上的差别，其数值必然存在差异。此外，该方法直接得到的不同年龄的预期寿命，其已存活年龄与现有方法不同，要想得到现有方法给出整数年龄的预期寿命，可以根据该结果，采用插值法估算。

本方法需要的的数据很少，仅需要两种人口统计数据，与现有统计部门公布的数据略有差异，就是需要统计的分年龄死亡人口应依据统计时间计算的年龄来分组，而不是依据死亡时的年龄来分组，就可得到需要的数据，这并不需要增加工作量，仅仅需要改变统计依据。通常人口普查或抽查，都获得死亡人口的出生时间数据，只要改变统计依据，根据出生时间和普查或抽查标准时间计算死亡者年龄，并进行统计，就可以获得本文方法所要求的死亡数据。本文所述方法计算得到的生命表各项数据都可依据分年龄人口和分年龄死亡人口统计数据直接计算得到的，不需要引入近似，所得生命表各项数据都是准确反映统计数据。

本文所述完整生命表编制方法，很简单易学，不容易出错，而且比传统方法准确可靠，在人口学领域应该具有较大的应用价值。